\documentstyle[12pt,aasms4]{article}   

\lefthead{S.\ Cassisi et al.}   

\righthead{$\alpha-$enhanced color transformations}   

\begin{document}   
\title{Color transformations and bolometric corrections
for Galactic halo stars: $\alpha-$enhanced {\sl vs} scaled-solar results.}   
   
\author{Santi Cassisi\altaffilmark{1,2}, Maurizio Salaris\altaffilmark{3},
Fiorella Castelli\altaffilmark{4,5}, and  Adriano Pietrinferni\altaffilmark{1,6}}   
   
\altaffiltext{1}{INAF - Osservatorio Astronomico di Collurania, Via M.\ Maggini,    
I-64100 Teramo, Italy; cassisi,pietrinferni@te.astro.it}   
      
\altaffiltext{2}{Instituto de Astrofisica de Canarias, 38200 La Laguna, Tenerife, Canary
Islands, Spain}

\altaffiltext{3}{Astrophysics Research Institute, Liverpool John Moores University,   
Twelve Quays House, Egerton Wharf, Birkenhead CH41 1LD, UK; ms@astro.livjm.ac.uk}

\altaffiltext{4}{INAF-Osservatorio Astronomico di Trieste, Via
G.B. Tiepolo 11, Trieste, Italy; castelli@ts.astro.it}

\altaffiltext{5}{Istituto di Astrofisica Spaziale e Fisica 
Cosmica, CNR, Via del Fosso del Cavaliere, 00133, Roma, Italy}

\altaffiltext{6}{Dipartimento di Statistica, Universit\`a di Teramo, Loc. Coste
S. Agostino, 64100 Teramo, Italy}

\begin{abstract}   
We have performed the first extensive analysis of the impact of an 
[$\alpha$/Fe]$>$0 metal
distribution on broadband colors in the 
parameter space (surface gravity, effective temperature, metal
content) covered by Galactic globular cluster stars.
A comparison of updated and homogeneous 
ATLAS~9 $UBVRIJHKL$ synthetic photometry, for
both $\alpha-$enhanced and scaled-solar metal distributions, 
has shown that it is impossible to reproduce
$\alpha-$enhanced $(B-V)$ and $(U-B)$ color transformations 
with simple rescalings of the scaled-solar ones. 
At [Fe/H]$\sim -$2.0 $\alpha-$enhanced transformations are well
reproduced by scaled-solar ones with the same [Fe/H], but this
good agreement breaks down at [Fe/H] larger than about $-$1.6.
As a general rule, $(B-V)$ and $(U-B)$ $\alpha-$enhanced colors 
are bluer than scaled-solar ones at either the same [Fe/H] 
or [M/H], and the differences increase with increasing metallicity and
decreasing $T_{eff}$. 
A preliminary analysis of the contribution of the various
$\alpha$ elements to the stellar colors shows that 
the magnesium abundance (and to lesser extent oxygen and silicon) 
is the main responsible for these differences.
On the contrary, the bolometric correction to
the $V$ band and more infrared
colors predicted by $\alpha-$enhanced transformations are well
reproduced by scaled-solar results, due to their weak dependence on
the metal content.
Key parameters like the Turn Off and Zero Age Horizontal Branch $V$ magnitudes, as
well as the red Giant Branch tip $I$ 
magnitude obtained from theoretical isochrones 
are in general unaffected when using the appropriate $\alpha-$enhanced
transformations in place of scaled-solar ones.
We have also studied for the first time the effect of boundary
conditions obtained from appropriate $\alpha$-enhanced model
atmospheres on the stellar evolutionary tracks in the
log$L/L_{\odot}$-$T_{eff}$ plane. We find that, for both scaled solar
and $\alpha$-enhanced metal mixtures, the integration of a 
solar $T(\tau)$ relationship
provides -- at least for masses larger than 0.5 -- 0.6 $M_{\odot}$ -- 
tracks very similar to the ones computed
using boundary conditions from the appropriate model atmospheres.
\end{abstract}   
   
   
\section{Introduction}   
   
Theoretical stellar evolution models, color transformations and
bolometric corrections are essential ingredients for interpreting
photometric data of resolved and unresolved stellar systems.
A reliable comparison of theory with observations requires, in
principle, that the element abundance distribution employed in the
theoretical models closely matches the pattern in the observed stellar
systems.

Whereas the standard heavy element distribution generally used in
stellar evolution computation is the solar one, there are 
cases where this assumption is not correct. A remarkable example is
the Galactic halo, whose stellar
component (field and globular cluster stars) shows a
ratio [$\alpha$/Fe]$>$0 (e.g. Lambert~1989) 
where $\alpha$ denotes elements
like O, Ne, Mg, Si, S, Ar, Ca and Ti.
The effect on stellar models and isochrones of a metal ratio 
[$\alpha$/Fe]$>$0 has been exhaustively investigated by
Salaris, Chieffi \& Straniero~(1993). They found that for
$\alpha$ element distributions typical of the halo
population, $\alpha-$enhanced models and isochrones are well reproduced by
scaled-solar ones with the same global metal abundance. This result
has been widely used until now, with the proviso that -- 
as shown by, e.g., Weiss, Peletier \& Matteucci~(1995), Salaris \& Weiss~(1998),
Vandenberg et al.~(2000), Salasnich et al.~(2000), 
Kim et al.~(2002) -- this equivalence breaks down when
the total metallicity is of the order of Z$\approx$0.002.

What is still missing is a study of the influence of [$\alpha$/Fe]$>$0
on the color transformations and bolometric corrections (hereafter CT
transformations) derived from theoretical model atmospheres. 
Until now all published sets of
$\alpha-$enhanced evolutionary models employ theoretical CT transformations 
derived from scaled-solar model atmosphere grids (sometimes with 
empirical adjustments), although it is not clear
to what extent scaled-solar transformations are a good approximation to the
proper $\alpha-$enhanced colors and bolometric corrections.

Barbuy~(1994), McQuitty et al.~(1994), Tripicco \& Bell~(1995),
Barbuy et al.~(2003), Thomas, Maraston \& Bender~(2003), Vazdekis et al.~(2004), 
Franchini et al.~(2004),
have investigated the effect of an $\alpha-$element enhancement on the
spectral indices used for metallicity and age estimates of
unresolved stellar systems, but there are no analogous studies 
devoted to the effect on broadband colors and bolometric corrections.
The effect of an $\alpha-$element enhancement on $(U-B)$ and $(B-V)$ 
colors of Main Sequence stars and red giants has been 
only briefly discussed by Castelli~(1999) for the case of a single low 
metallicity value, i.e. [Fe/H]=$-$2.

This paper aims at filling this gap, by studying the differences between
scaled-solar and $\alpha-$enhanced transformations to the 
widely used $UBVRIJHKL$ photometric bands, for Galactic halo stars.
In addition, our grid of $\alpha$-enhanced
model atmospheres allowed us to employ boundary
conditions for the stellar model computations obtained from the
appropriate model atmospheres, and compare 
in the log$L/L_{\odot}$-$T_{eff}$ plane the results with models
computed using -- as customary -- boundary conditions obtained 
by integrating a solar $T(\tau)$ relationship. 
In \S\ref{model} we briefly introduce the synthetic colors and
bolometric corrections used in this work,
and in \S\ref{comp} we compare scaled-solar with $\alpha-$enhanced CT
transformations. In \S\ref{bound} we discuss the effect of the 
boundary condition choice on the stellar model computation, 
and the full results are summarized in the final section.
      
\section{Model atmospheres, colors and bolometric corrections}\label{model}

$UBVRIJHKL$ synthetic photometry based on updated ATLAS~9 model
atmospheres (Castelli \& Kurucz~2003) was computed for both scaled-solar 
and $\alpha-$enhanced metal distributions (see Pietrinferni et al.~2004 for
more details). The model atmospheres were computed by adopting as reference
solar chemical composition that from Grevesse \& Sauval~(1998) instead
of the Anders \& Grevesse~(1989) one, as in the case of previous ATLAS9 models
(Kurucz 1993). Previous scaled solar models were computed for an iron abundance 
log(N$_{Fe}$)=7.67 (with the usual normalization log(N$_H$)=12), 
whereas $\alpha-$enhanced models were computed
assuming the more recent estimate log(N$_{Fe}$)=7.51. 
This discrepancy of the Fe abundance for the old models prevented 
in the past a rigorous comparison of $\alpha$-enhanced and scaled-solar colors.
Furthermore, the new model atmospheres include now (for both scaled-solar and
$\alpha$-enhanced metal mixtures) the important 
contribution to the total opacity of H$_{2}$O lines and of the 
quasi-molecular absorption of H-H and H-H$^+$ (Castelli \& Kurucz~2001).

The $\alpha-$enhanced models have scaled-solar abundances
for all elements except O, Ne, Mg, Si, S, Ar, Ca and Ti,
for which the logarithmic scaled-solar abundance is increased
by 0.4 dex (i.e., [$\alpha$/Fe]=0.4).  
All the models (both scaled-solar and $\alpha$-enhanced) 
were computed with the overshooting option for the convection switched off
and a mixing-length parameter l/H$_p$=1.25 (see Castelli~1999 and reference
therein). 
Grids of updated model atmospheres, energy distributions
and colour indices in $UBVRIJKL$ and Str\"omgren $uvby$ photometric systems
are available so far for [Fe/H] equal to $-$2.5, $-$2.0, $-$1.5,
$-$1.0, $-$0.5, 0.0 in case of the $\alpha$-enhanced metal distribution
(we are in the process of 
computing models for super-solar metal content) and 
extended up to [Fe/H]= +0.2 and +0.5  
for the scaled-solar mixture. In all the cases the 
microturbulent velocity is $\xi$=2~km~s$^{-1}$.
The adopted $U$ passband is from Buser (\cite{BU78}), $B$ and $V$ 
from Azusienis \& Strai\v zys (\cite{AS69}), $R$ and $I$ Cousins 
passbands from Bessell (\cite{B90}), $JKL$ 
passbands from Johnson (\cite{J65}) reported also by Lamla (\cite{L82}).
Finally, the $H$ passband is from Bessell \& Brett (\cite{BB88}). 
For each [Fe/H] value the model grid covers the range from 3500 K to 50000 K 
in $T_{eff}$, and from 0.0 to 5.0 in log($g$). These new grids 
(labeled ODFNEW grids) can be downloaded from http://kurucz.harvard.edu/grids.html
and from http://wwwuser.oat.ts.astro.it/castelli/grids.html.

\section{$\alpha-$enhanced transformations {\sl versus} scaled-solar
ones}\label{comp}

The complete consistency between the scaled-solar and
$\alpha-$enhanced CT transformations enables us to perform a
reliable differential comparison of the two sets.
More in detail, we have 
investigated the differences between our scaled-solar 
and $\alpha-$enhanced CT transformations
by comparing the Color-Magnitude-Diagrams (CMDs) of 
theoretical isochrones transformed from the log$L/L_{\odot}$-$T_{eff}$
plane using both sets of transformations, in different [Fe/H] regimes. 
The underlying isochrones we employed 
are $\alpha-$enhanced isochrones ($<$[$\alpha$/Fe]$>$=0.4) 
with a metal distribution very
similar to the one adopted in the model atmosphere calculations; the
stellar evolution code and input physics 
are the same as in Pietrinferni et al.~(2004). 
A subset of these
models has been already discussed in Cassisi, Salaris \& Irwin~(2003)
and Salaris et al.~(2004) 
where a concise summary of the adopted input physics can be found.
We wish to emphasize that here 
we are comparing differentially our two sets of CT transformations,
using the same underlying isochron.
The differences we find in the CMD location of the transformed
isochrone are therefore due only to differences in the 
transformations.
In principle we could have compared the CT transformation tables
on a point by point basis, but with our approach we are automatically
making the comparison within 
a parameter space typical of stars populating the
Galactic halo.

The most important outcomes of this analysis are the qualitative
differences between the CMDs obtained employing the two sets of
transformations. Since until now all results based on 
theoretical $\alpha$-enhanced isochrones have been obtained by
employing in principle inappropriate scaled-solar CT transformations, our analysis 
will allow to establish what results can be trusted and the ones that cannot.
These qualitative differences are not affected 
at all when we change the underlying theoretical isochrones 
(we obtain very similar results
using the Vandenberg et al.~2000 $\alpha$-enhanced isochrones),
as long as they reasonably approximate the [Fe/H], surface gravity and
effective temperature range of halo stars.
However, the precise numerical values of the differences can be
slightly dependent on the selected theoretical isochrone.

\begin{figure}[t]
\plotone{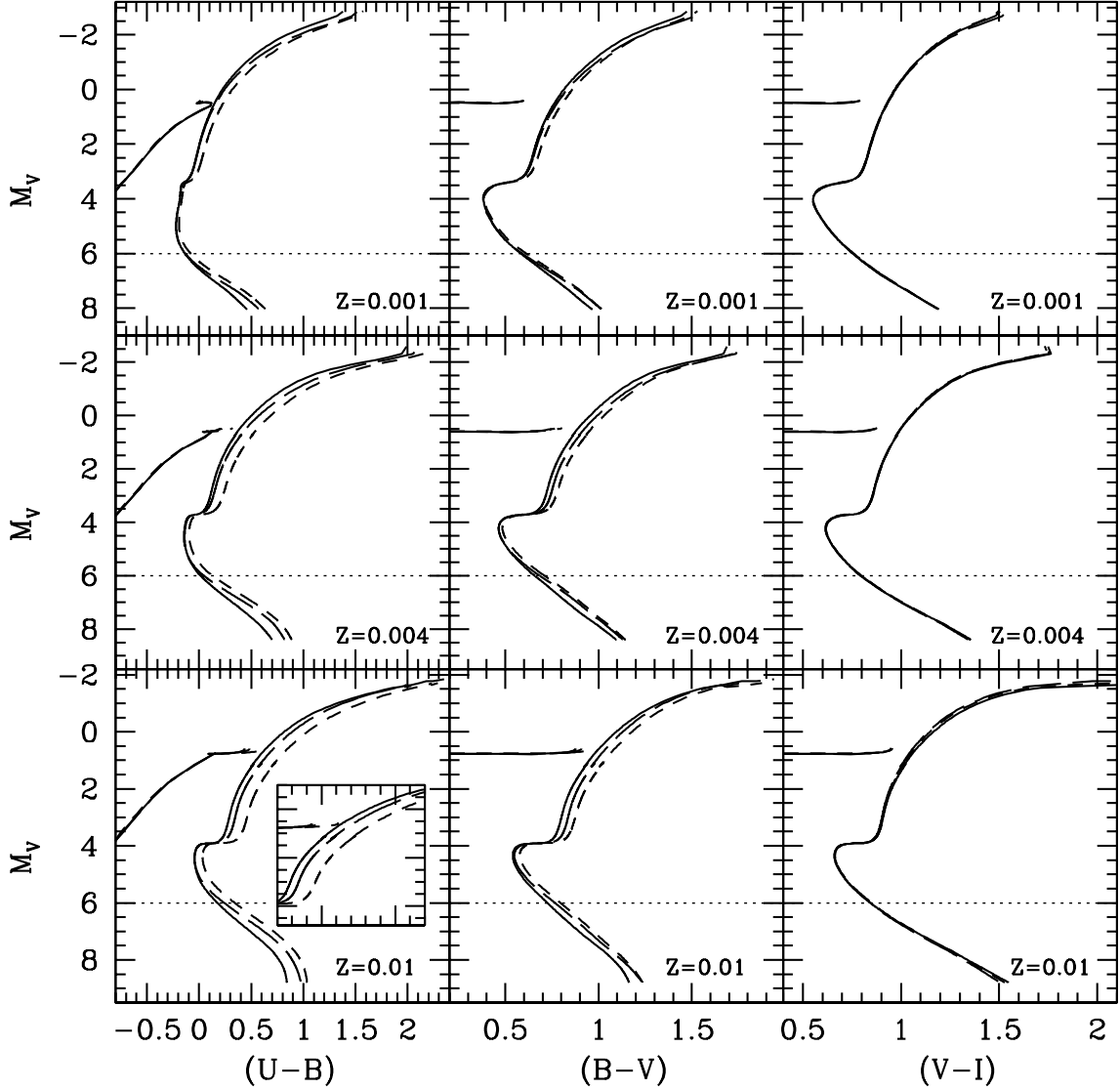}   
\caption{CMDs of 12~Gyr old $\alpha-$enhanced isochrones with the
labeled metallicities. Three different sets of CT transformations
have been employed. Solid lines represent $\alpha-$enhanced
transformations with the appropriate metal mixture; long dashed lines
denote scaled-solar transformations with the same [Fe/H] of the
$\alpha-$enhanced ones, whereas short dashed lines display the case of
scaled-solar transformations with the same [M/H] of the
$\alpha-$enhanced ones. 
The inset shows an enlargement of the RGB
section of the Z=0.01 isochrone, to enhance the difference
between the location of the solid and long dashed lines.
The horizontal dotted line marks the $M_V=6$ level (see text for details).  
\label{figiso}}   
\end{figure}

Figure~\ref{figiso} displays a theoretical isochrone 
and the corresponding Zero Age Horizontal Branch (ZAHB) 
for an age of  12~Gyr -- taken as representative of the 
typical Galactic globular cluster age (see, e.g., Salaris \&
Weiss~1998, 2002) -- and the following three
metallicities and helium abundances: Z=0.001, Y=0.246 -- Z=0.004,
Y=0.251 -- Z=0.01, Y=0.259.
These compositions, coupled with [$\alpha$/Fe]=0.4 
correspond to [M/H]=$-$1.27, [Fe/H]=$-$1.57 ([M/H] is the global
metallicity in spectroscopical notation) -- 
[M/H]=$-$0.66, [Fe/H]=$-$0.96 -- [M/H]=$-$0.25, [Fe/H]=$-$0.55.
We briefly notice here that the relationship between [Fe/H] and 
[M/H] for the scaled-solar
Grevesse \& Sauval~(1998) metal distribution and [$\alpha$/Fe] values 
typical of the Galactic halo, is well approximated by the following
relationship given by Salaris et al.~(1993):
[M/H]=[Fe/H]+log(0.638\ f\ +0.362), where log(f)=[$\alpha$/Fe].

We have first transformed the three isochrones using the $\alpha-$enhanced
transformations (solid lines in Fig.~\ref{figiso}) for the appropriate 
[Fe/H] (hence [M/H]). Then we used scaled-solar transformations, as routinely done
in the literature, although it is not usually specified how they are
applied to $\alpha-$enhanced models. The issue is that,  
when employing scaled-solar transformations, one has to appropriately 
choose the independent variable for the interpolation among
the CT transformation tables
(the same choice has to be made when comparing the corresponding
transformation tables).
%
There are two simple and natural choices for this. The first one
is to consider the total metallicity [M/H] of the $\alpha-$enhanced
models, and determine the scaled-solar transformations at the same
[M/H]. This means that the individual metal abundances used in the
transformations are different from the models, but the global metal
content is the same; in our case, the Fe abundance for the scaled
solar synthetic colors would be about 0.3 dex higher than the proper
one. Our isochrones transformed in this way 
are displayed as short dashed lines in Fig.~\ref{figiso}.
The second possibility is to consider scaled-solar transformations with
the same [Fe/H] of the $\alpha-$enhanced models; this choice assumes
that it is the Fe abundance (and eventually the elements that are not
enhanced with respect to Fe) that contributes mostly to the observed
colors and bolometric corrections. This also means that the
transformations would take into account the appropriate abundance of Fe
and other scaled-solar elements, while underestimating the abundance of
the $\alpha-$elements by 0.4 dex in our case. 
Isochrones transformed in this way are shown as
long dashed lines in Fig.~\ref{figiso}.

Figure~\ref{figiso} shows clearly that the (M$_{V}, V-I$) CMD, regardless of the
cluster metallicity, is unaffected by the CT choice (the same is true
for metallicities below Z=0.001); this holds 
also for (M$_{V}, V-R$) and other near infrared CMDs included in our set of
transformations. This result is hardly surprising, since in general 
$(V-R)$, $(V-I)$, $(V-K)$, $(J-K)$ and $(H-K)$ color transformations are weakly 
sensitive to the metal content. Also $BC_V$ 
values appear to be very weakly affected by the
selected transformations.
 
The case of bluer colors like $(B-V)$ and $(U-B)$ is quite different.
We consider first the comparison between the proper
$\alpha-$enhanced transformations and the scaled-solar ones computed
for the same [M/H]. 
At Z=0.001 ([Fe/H]=$-$1.6) there are discrepancies 
of a few hundredths of magnitudes in the colors (which are reduced at
lower metallicities to within $\sim$0.01 mag for [Fe/H]=$-$2);  
these differences 
increase with increasing metallicity, and are larger in $(U-B)$ than 
in $(B-V)$.
In general, scaled-solar color transformations selected on the
basis of the isochrone [M/H] do produce redder $(B-V)$ and $(U-B)$ values.
The color shift between the two sets of transformations is also a 
function of the effective temperature; it generally increases for decreasing
$T_{eff}$, as clearly seen in Fig.~\ref{figiso}. The global effect is
therefore also a slight change of the isochrone morphology in the 
(M$_{V}, B-V$) and (M$_{V}, U-B$) CMDs. 
However, since the $UB$ photometry is employed mainly for 
studying hot HB stars, which are hot enough to be unaffected by the choice
of the transformations, the differences are not extremely relevant 
when using this color index \footnote{There exists a third possible way of 
mimicking the CT transformations for $\alpha-$enhanced
mixtures, that is to use scaled-solar transformations computed for 
the same [$\alpha$/H] abundance of the $\alpha-$enhanced models. 
In this way one assumes that mainly the $\alpha-$elements affect the
transformations; the chosen metal content for the scaled-solar
transformations would have in this case a Fe
abundance 0.4~dex higher than the proper $\alpha-$enhanced
distribution. The data plotted in Fig.~\ref{figiso}, clearly show that 
this choice would produce the worst results, since it would correspond to a scaled-solar [M/H]
value 0.4 dex higher than the actual one, hence yielding even redder 
$(B-V)$ and $(U-B)$ colors.}.
%
%

\begin{figure}[t]
\plotone{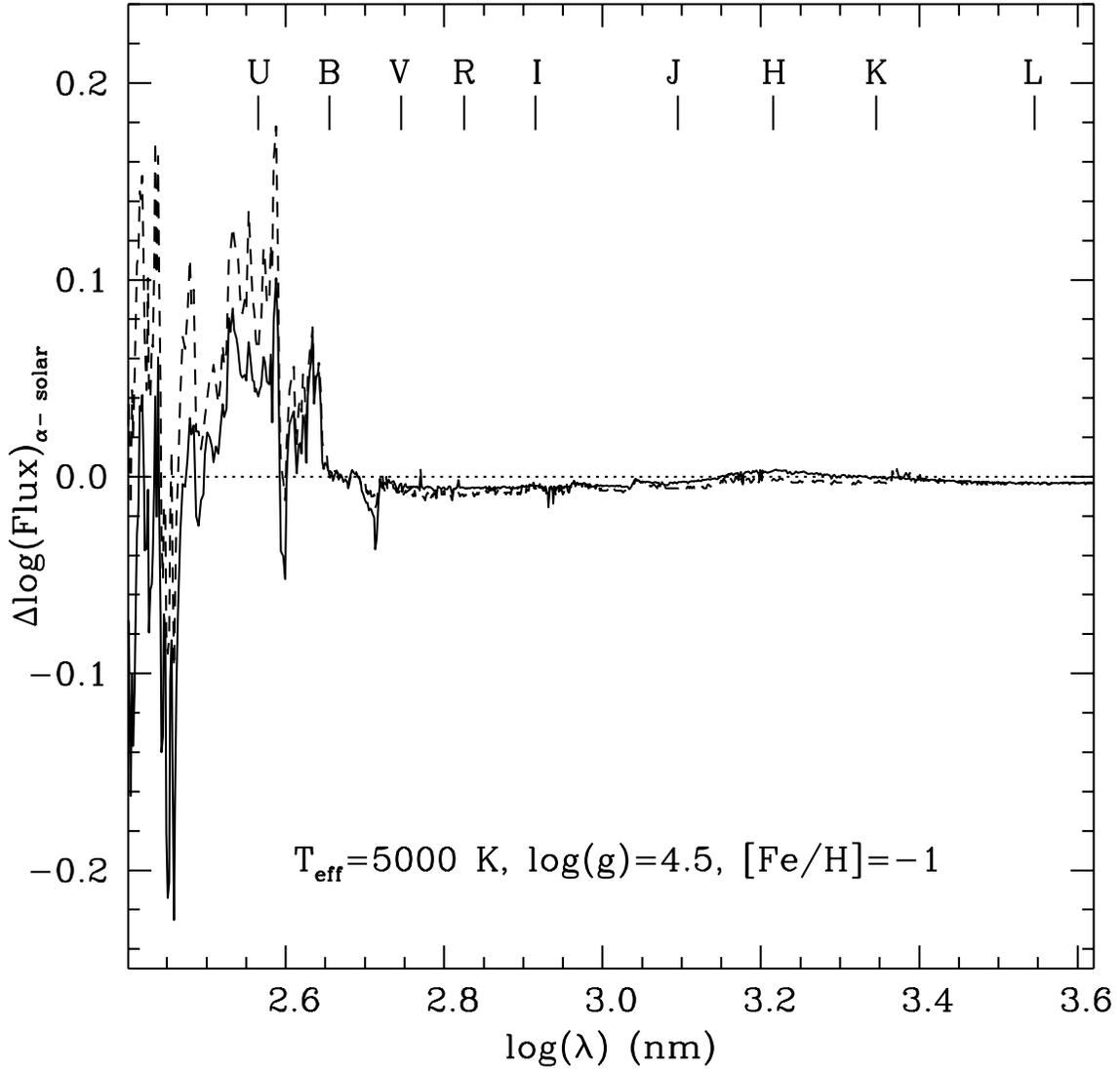}   
\caption{Difference of the logarithm of the flux (erg/{\rm
$cm^2$}/s/hz/ster) predicted for an $\alpha-$enhanced model 
$T_{eff}$=5000~K, log($g$)=4.5, [Fe/H]=$-$1.0 ([M/H]=$-$0.7), 
and a corresponding scaled-solar model with either the same [Fe/H] (solid
line) or the same [M/H] (dashed line). 
Wavelengths are in nm; the effective wavelengths of various 
broadband filters are also marked.
\label{figflux}}   
\end{figure}

When considering the case of scaled-solar transformations with the
same [Fe/H] of the $\alpha-$enhanced mixture, 
the results are practically identical at [Fe/H] of the order of $-$2.0
and in general one obtains a better
agreement with the appropriate $\alpha-$enhanced transformations;
however, the discrepancies are non negligible when Z$\sim$0.001, 
increasing with both metal content and decreasing $T_{eff}$. 
At first it may appear surprising that $(B-V)$
and $(U-B)$ colors at the same [Fe/H] are bluer in the
$\alpha-$enhanced case, when the total metal content is higher. 
Figure~\ref{figflux} shows the differences between the flux predicted 
for a typical $\alpha-$enhanced MS model with [Fe/H]=$-$1, and 
scaled-solar models with either 
the same [Fe/H] (solid line) or the same [M/H] (dashed line).
The $\alpha-$enhanced model
shows a larger flux in the $UB$ part of the spectrum with respect to
the scaled-solar counterpart with the same [Fe/H], and a qualitative
analysis of the figure shows clearly that 
the $\alpha-$enhanced $(U-B)$ and $(B-V)$ indices have to be bluer. 
On the other hand, colors built from passbands bluer than $U$
may provide different results. It is in fact evident from
Fig.~\ref{figflux} that, e.g., around $\lambda$=280 nm 
(log($\lambda$)$\sim$2.45) the $\alpha-$enhanced
flux is definitely lower than the scaled-solar counterparts.

In order to explain the differences highlighted by 
Fig.~\ref{figflux}, one needs to assess the role played by the different $\alpha-$elements 
in modifying the stellar flux with respect to a scaled-solar mixture.
It is clear that, for an exhaustive analysis, one should compute 
a set of model atmospheres for several
gravities and effective temperatures, by changing the abundance of each
individual $\alpha$ element at constant [Fe/H]. 
This notwithstanding --  prompted by the referee
-- we have obtained relevant preliminary results 
by computing selected models for the same
[Fe/H], gravity and $T_{eff}$ as in Fig.~\ref{figflux}, 
enhancing once at a time the abundances 
of the most relevant $\alpha-$elements, i.e. O, Si, Mg and Ca.

As a general point, the different metal distributions of the $\alpha-$enhanced and scaled-solar
models, computed for the same $T_{eff}$ and gravity, affect both the line opacity and the 
continuum opacity, redistributing the flux  among the various
wavelengths in such a way that the flux conservation 
$\int_{0}^{\infty}{F_\nu d\nu}=\sigma T_{eff}^{4}/\pi$ is met.
For any abundance change 
the shape of the flux distribution is modified, due to a change of
either the continuum level,
or the line absorption or both. 
For instance, the $\alpha$ element that mostly affects the level of the continuum in 
our model is Mg, owing to the occurrence of a Mg I ionization edge 
at 2500$\AA$. By increasing the Mg abundance the continuum flux
decreases shortward of the Mg I discontinuity and increases longward of it.  
Instead, the oxygen abundance affects the flux distribution in the region of the U and V
bands through the numerous OH lines lying mostly shortward of the Balmer discontinuity.

%

\begin{figure}[t]
\plotone{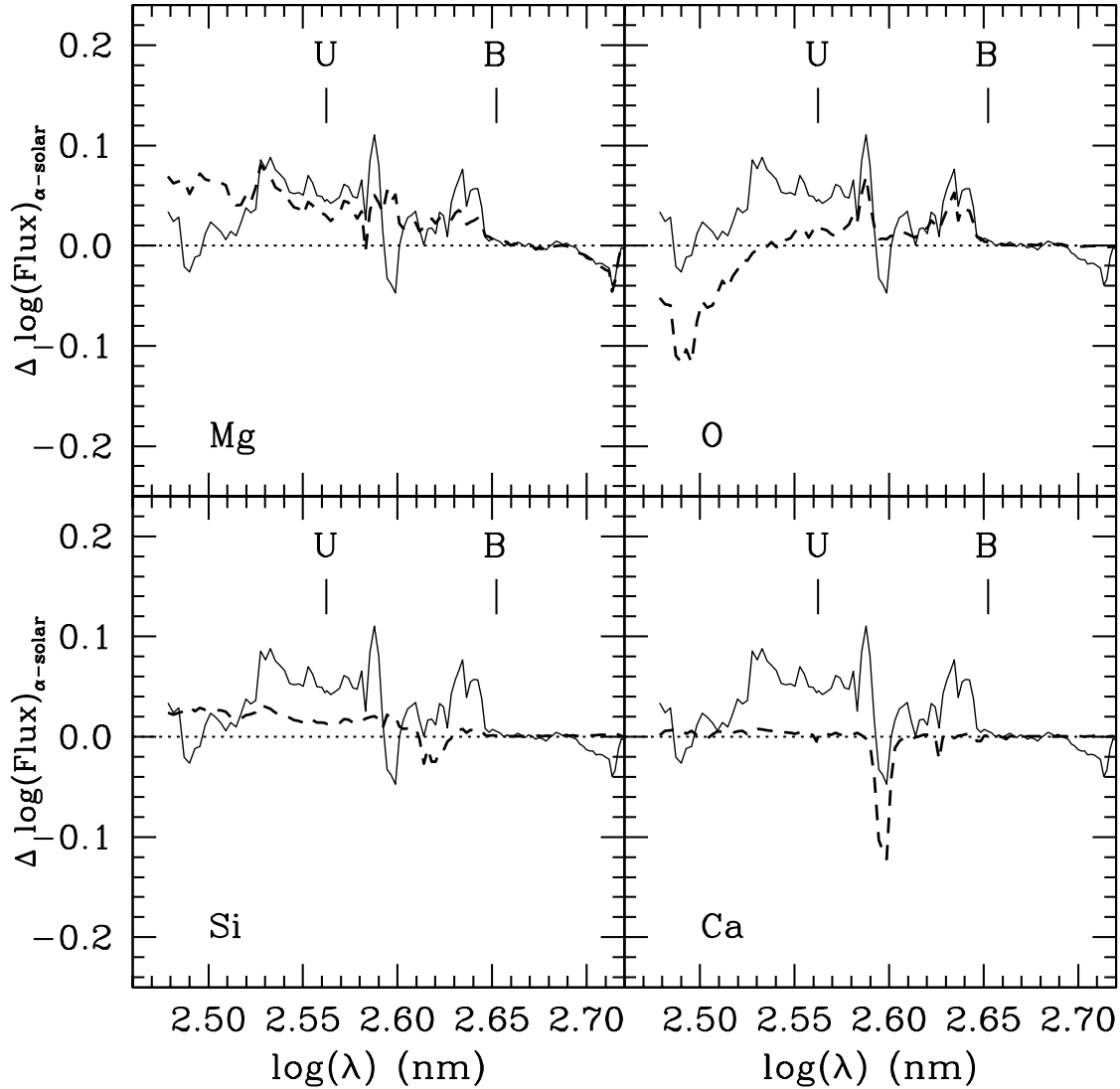}   
\caption{As in Fig.~\ref{figflux} but for 4 partial 
$\alpha$-enhanced mixtures where only the labeled elements 
are enhanced with respect to Fe ([$\alpha$/Fe]=0.4).
All mixtures have [Fe/H]=$-$1.0; the solid line displays the difference
between the complete $\alpha$-enhanced distribution and the solar one 
(it corresponds to the solid line in  Fig.~\ref{figflux}), while 
dashed lines show the differences between the various partial $\alpha$-enhanced
mixtures and the solar one. 
\label{figelement}}   
\end{figure}

Figure~\ref{figelement} shows 
differences $\Delta$(Flux) between the flux predicted for models with 
only the labeled $\alpha$ element enhanced (by 0.4 dex each), and the scaled solar
case with the same [Fe/H]=$-$1.0, $T_{eff}=5000K$ and $\log(g)=4.5$.
The
$\Delta$(Flux) values between the complete $\alpha$-enhanced model and the scaled
solar one with the same [Fe/H], $T_{eff}$ and $\log(g)$ is also displayed.
We focused our analysis on the 
wavelength region where the largest $\Delta$(Flux) values contributing to the
discrepancies in the $U$ and $B$ filters are found. 

One can notice how $\Delta$(Flux) between the complete $\alpha$-enhanced
model and the scaled solar one is essentially due to just O, Si, Mg and Ca.
Calcium contributes mainly to the absorption at 
$\lambda \sim$390 nm, whereas O and Mg play a major role in the
differences at $\lambda > 350$ nm (apart from the Ca absorption
mentioned before). In addition, Si and Mg are the most important contributors
to the region shortward of $\sim$350 nm. 

Table~1 displays the $(U-B)$ and $(B-V)$ colors predicted by the same 
model atmospheres whose flux difference are shown in
Fig.~\ref{figelement}; this allows us to assess quantitatively 
how much these individual $\alpha$ elements affect the broadband $UBV$
colors. Magnesium appears to be the main responsible for the 
$(B-V)$ and $(U-B)$ color
differences between the scaled solar mixture and the full 
$\alpha-$enhanced one; in fact, by enhancing only Mg 
one recovers 88\% and 76\% of the differences in $(B-V)$ and $(U-B)$, respectively.
Oxygen has a smaller but non negligible impact
on the $(B-V)$ color whereas the $(U-B)$ color is affected appreciably
also by silicon. Calcium and the other $\alpha$ elements have only a
minor impact on these two colors at this $T_{eff}$ and gravity.

Prompted by our referee, to investigate the effects due to surface gravity 
changes along an isochrone, we have computed additional model atmospheres by enhancing the abundance
of each individual $\alpha-$element, at the same [Fe/H] and effective
temperature as above, but for $\log(g)=3.0$. 
We obtain results similar to the case with  $\log(g)=4.5$ 
-- as one could also expect by comparing the CT tables for scaled solar
and full $\alpha$-enhanced mixtures at these two gravities -- 
the main difference being that, whereas at $\log(g)=4.5$ the most important
$\alpha-$element for the $UB$ flux distribution is Mg, in this case the contributions provided by Mg
and Si are comparable.

%
%

Table~2 lists the differences between some
relevant features of the three sets of isochrones
displayed in Fig.~\ref{figiso}. In particular, we wish to briefly 
discuss the $(B-V)$ color
of the Main Sequence (MS) at M$_{V}$=6
($(B-V)_{\rm M_{\sl V}=6}$), the absolute visual magnitude of
the Turn Off (TO), the color
difference $(\Delta(B-V))$ between the TO and the Red
Giant Branch (RGB), 
the absolute I-Cousins magnitude of the RGB tip and the absolute visual
magnitude of the ZAHB at the level of the instability strip.
These quantities are usually employed in studies about 
distance and age determinations of old stellar populations.
In the following we will refer to the use of scaled-solar 
transformations with the same [M/H] of the $\alpha-$enhanced mixture,
which produce the larger differences.

The $(B-V)_{\rm M_{\sl V}=6}$ color (taken as representative magnitude of the
unevolved MS in globular clusters) enters the 
MS fitting method for deriving cluster distances 
(e.g., Carretta et al.~2000, Percival et al.~2002).
In fact, the color shifts applied to individual subdwarfs in
order to \lq{register}\rq\ their own metallicity to that of the
considered cluster, are usually derived from theoretical isochrones, at
least in the regime of globular cluster metallicities. The
data listed in Table~2 show that the $(B-V)_{\rm M_{\sl V}=6}$
difference between the metallicities displayed in
Fig.~\ref{figiso} does depend on the choice of the transformations, 
and are about 20\%
smaller when using the appropriate $\alpha-$enhanced transformations.
This discrepancy may however have only a small impact on the derived distances
as long as the subdwarfs employed in the MS-fitting have metallicities
close to the cluster one, hence the color shifts applied are small.

The $(B-V)$ color extension between TO and RGB, $\Delta(B-V)$, is 
also clearly strongly affected. Following Rosenberg et al. (1999), 
here we have defined $\Delta(B-V)$ as the color difference between
the TO and a point on the RGB 2.5~mag brighter than the TO.
This quantity is a function of the cluster
age, but it is generally used only in a differential way in order to
determine relative cluster ages (e.g. Vandenberg, Bolte
\& Stetson~1990, Salaris \& Weiss~1998, Rosenberg et al.~1999).
The absolute values of $\Delta(B-V)$ change by $\sim$ 0.03-0.04 mag,
depending on the metallicity; this would cause an age 
variation by $\sim$ 2-3 Gyr if $\Delta(B-V)$ is
used to estimate absolute ages. When differences of 
$\Delta(B-V)$ at varying ages are used for relative age estimates, the
impact of the CT transformation choice is however almost negligible.

As for the M$_{V}$ values of TO and ZAHB at the RR Lyrae instability
strip -- whose difference is usually employed to estimate cluster ages
-- they are affected by the adopted CT relation at the level of $\approx$ 0.01 mag,
which has a negligible impact on the age estimates.
We have in addition verified that, at a fixed metallicity, the range 
of masses populating the
RR~Lyrae instability strip is basically unaffected by the choice of the
transformations. 
Also the $I$ magnitude of the RGB Tip (a widely used distance
indicator for old stellar populations, see e.g. Lee, Freedman \&
Madore~1993, Salaris \& Cassisi~1998 and references therein) is 
unaffected by the choice of the transformations, 
the variation being of the order of $\sim$0.01 mag, which has a
negligible impact on the distance determinations.

\section{Treatment of surface boundary conditions}\label{bound}

It is well known that in order to integrate the stellar structure
equations, it is necessary to fix the value of the pressure and
temperature at the stellar surface, usually close to the photosphere.
There are basically two possibilities to determine this value. The
first one is to integrate the atmospheric layers by using a $T(\tau)$
relationship, supplemented by the hydrostatic equilibrium condition
and the equation of state; the second possibility is to obtain the
required boundary conditions from precomputed non-gray model
atmospheres.

The first procedure is universally used in stellar model
computation; i.e., in our stellar evolution 
calculations employed above we have used the
Krishna-Swamy~(1966) solar $T(\tau)$ relationship.
In Salaris, Cassisi \& Weiss~(2002) we have already shown that 
in case of scaled-solar models, the $T(\tau)$ integration and boundary
conditions from model atmospheres (belonging to a previous ATLAS~9 release)
provide RGB tracks that agree within about 50~K.

\begin{figure}[t]
\plotone{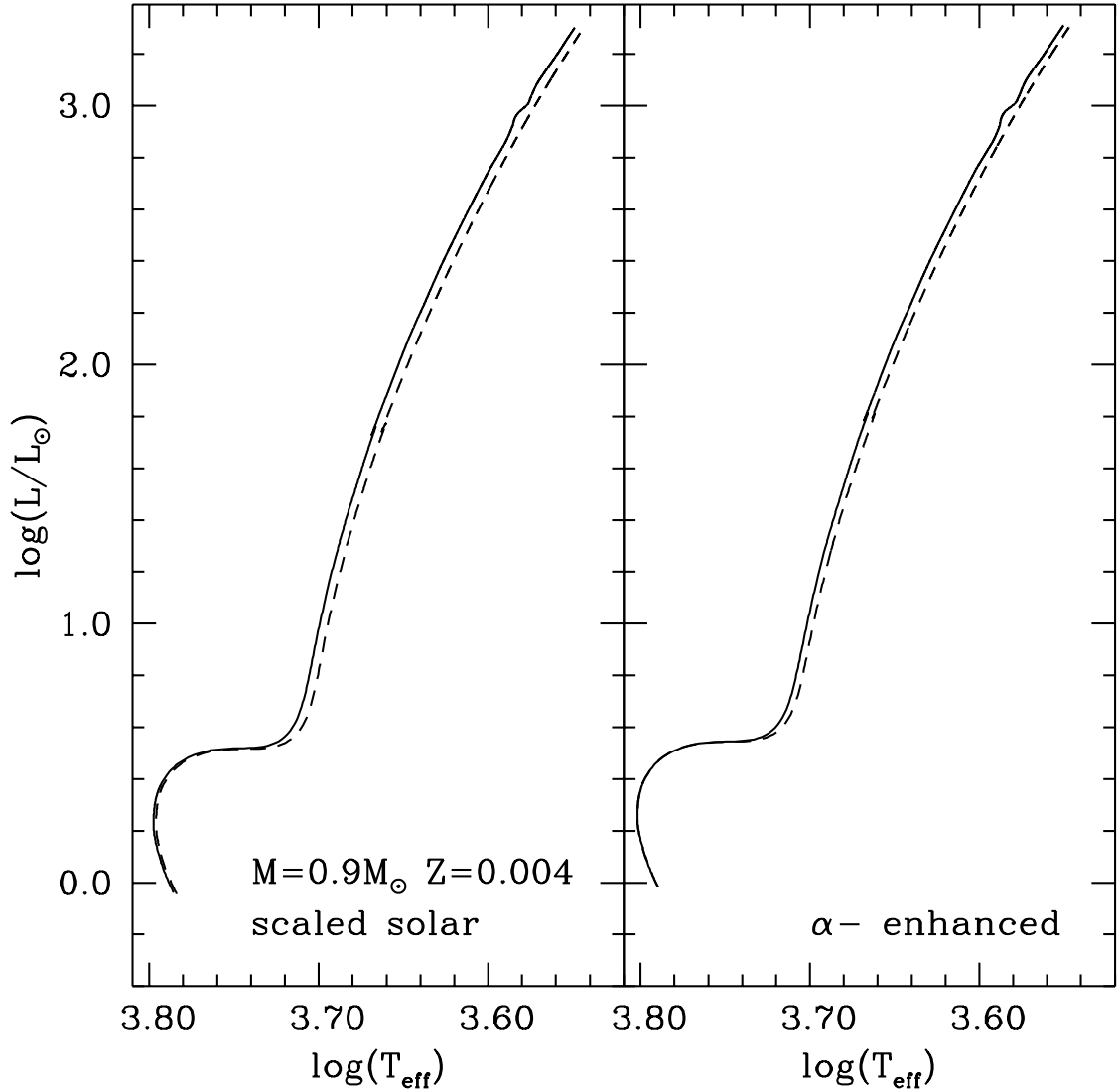}   
\caption{Comparison on the log$L/L_{\odot}$-$T_{eff}$ plane of
0.9$M_{\odot}$ stellar models with Z=0.004 (scaled-solar and
$\alpha$-enhanced) computed using boundary conditions from model
atmospheres (dashed lines) and from the integration of a $T(\tau)$
relationship (solid line). 
\label{figboundcon}}   
\end{figure}

Here we repeat the test on scaled-solar models and for the first time we add a
corresponding test for an $\alpha$-enhanced mixture,
using the updated ATLAS~9 model grid discussed in the previous section.
Figure~\ref{figboundcon} shows the evolutionary tracks of a
0.9$M_{\odot}$ star with a turn off age of $\sim$ 10.5 Gyr, from the
beginning of the MS up to the RGB tip, in the log$L/L_{\odot}$-$T_{eff}$ plane. 
Four different tracks are
displayed, corresponding to the pair Z=0.004, Y=0.251, for scaled
solar and $\alpha$-enhanced metal distributions, computed
using boundary conditions from both a $T(\tau)$ integration and 
non-gray model atmospheres with the appropriate metal mixture.
The boundary conditions from the model atmospheres were taken at 
$\tau$=56.
%
%
In both scaled-solar and $\alpha$-enhanced case the MS is completely
insensitive to the choice of the boundary conditions. The RGB part is 
slightly affected, at the level of at most 40~K, the model
atmosphere tracks being cooler. The effect of this temperature
change on the predicted colors is the following: $\Delta(B-V)\approx+0.02$ mag, 
$\Delta(U-B)\approx+0.03$ mag, $\Delta(V-I)\approx+0.02$ mag.

Evolutionary timescales and interior
properties of the models are also unaffected by the choice of the
boundary conditions.
Analogous results have been obtained at different Z. We also computed
a model for $M=0.9M_{\odot}$, Z=0.004, 
by taking the model atmosphere boundary conditions at
$\tau$=10, obtaining the same results as for the $\tau$=56 case.

We therefore conclude that scaled-solar
and $\alpha$-enhanced isochrones 
can be safely computed - within the quoted uncertainty of about 40~K -
integrating a solar $T(\tau)$ relationship 
for the boundary conditions, at least when the evolving mass is 
larger than $\sim 0.5-0.6 M_{\odot}$ (which is the lower mass limit  
of our isochrones); lower masses may be more affected
by the choice of the boundary conditions, as discussed, e.g., in
Alexander et al.~(1997), Chabrier \& Baraffe~(1997) and references therein.  
Moreover the results of all previously published  
comparisons between scaled-solar and
$\alpha$-enhanced models in the log$L/L_{\odot}$-$T_{eff}$ plane 
(e.g., Salaris et al.~1993, Vandenberg et
al.~2000) that were computed employing $T(\tau)$-based boundary
conditions, are fully confirmed when employing the boundary conditions
from the appropriate non-gray model atmospheres.

Our results suggest that employing a solar $T(\tau)$ relationship
for the boundary conditions provides a fair approximation to the 
boundary conditions from non-gray model atmospheres, although in
principle it is more appropriate and self-consistent to rely on 
model atmosphere results. This notwithstanding, one has to consider
the fact that also model atmospheres are affected by intrinsic
uncertainties, especially related to convection, and are usually based
on a convection treatment different from the one adopted in stellar
evolution computations (Montalban et al. 2001).

\begin{figure}[t]
\plotone{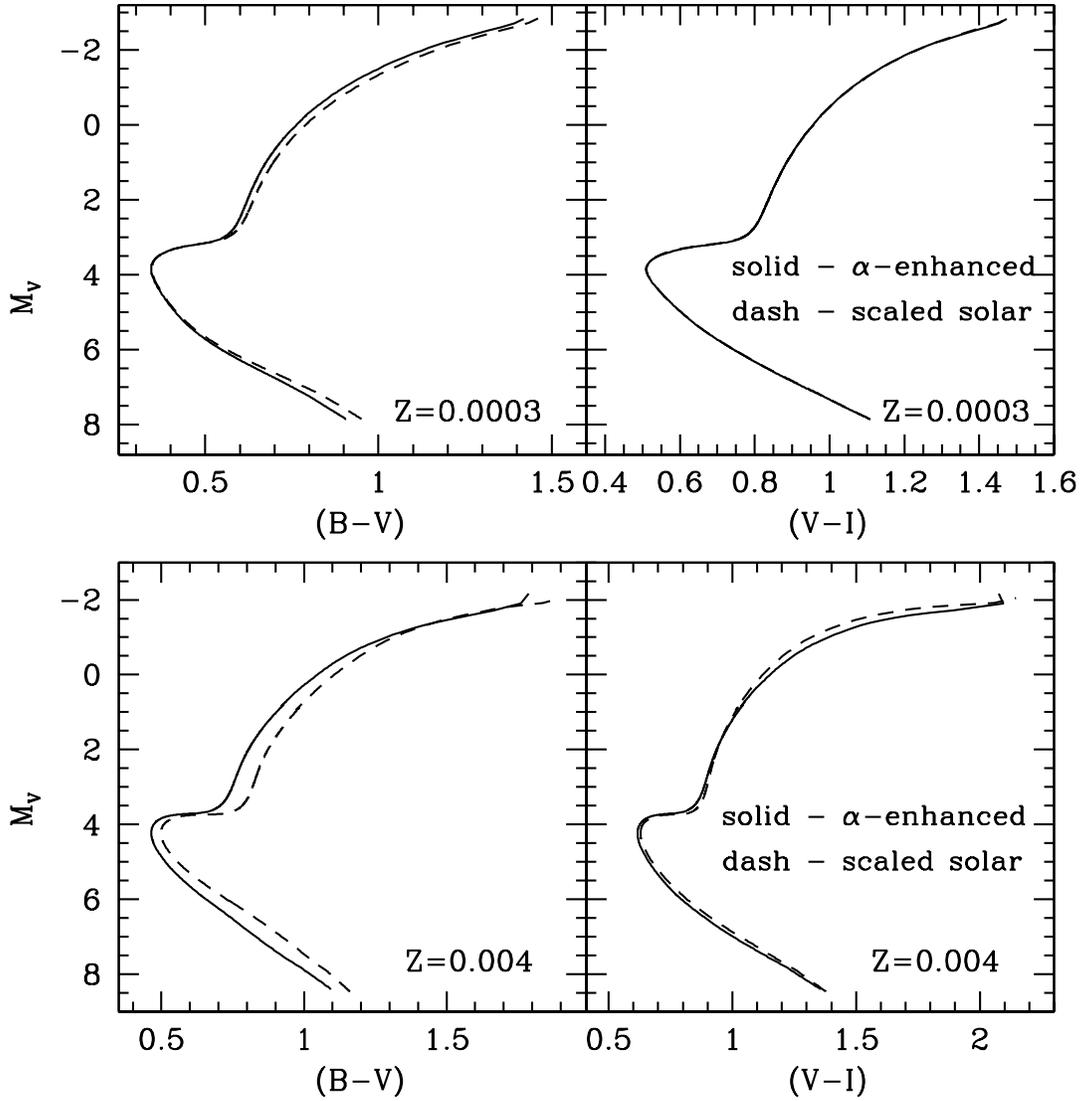}   
\caption{CMDs of 12~Gyr old isochrones computed by adopting a scaled solar mixture (dashed lines) and
an $\alpha-$enhanced one (solid lines) but the same global metallicity (the adopted metallicities are labeled
in the panels). For each adopted heavy element mixture, the boundary conditions in the stellar evolutionary computations
as well as the CT transformation have been consistently derived by an appropriate set of model atmospheres.
\label{figisobcct}}   
\end{figure}

Figure~\ref{figisobcct} compares 12~Gyr old 
isochrones for a scaled solar mixture and for 
an $\alpha-$enhanced one with the same global metallicity, computed by
using both boundary conditions and CT 
transformations from the appropriate model atmospheres. 
The equivalence 
between scaled-solar and $\alpha-$enhanced isochrones with the same
[M/H] is still good at low metallicities, especially in the $VI$
plane, while in the $BV$ plane there are small differences due to
the effect of the CT transformations. Larger differences are present 
at $Z\approx$0.004, especially significant in the $BV$ plane.

%
%
%
%
\section{Summary}\label{final}   

We have compared updated ATLAS~9 $UBVRIJHKL$ synthetic photometry for
both $\alpha-$enhanced and scaled-solar metal distributions, 
in a large range of metallicities typical of the Galactic halo populations. 
This is the first complete analysis of the impact of an [$\alpha$/Fe]$>$0 metal
distribution on broadband colors and bolometric corrections, for the 
full metallicity range of the Galactic halo population. 

We found that it is impossible to mimic the appropriate
$\alpha-$enhanced $(B-V)$ and $(U-B)$ color transformations 
with simple rescalings of the scaled-solar ones, 
over the entire [Fe/H] range of the Galactic halo. 
At [Fe/H]$\sim -$2.0 $\alpha-$enhanced transformations are well
reproduced by scaled-solar ones with the same [Fe/H], however, this
good agreement breaks down for [Fe/H] larger than about $-$1.6.
In general, $(B-V)$ and $(U-B)$ $\alpha-$enhanced colors 
tend to be bluer than scaled-solar ones at either the same [Fe/H] 
or [M/H], and the differences increase with increasing metallicity and
decreasing $T_{eff}$. 
These differences are mainly due to the enhancement of Mg 
with contributions from the enhancement of Si and O.

On the other hand $BC_{V}$ and more infrared
colors predicted by $\alpha-$enhanced transformations are well
reproduced by scaled-solar results.
Key quantities 
like the TO and ZAHB $V$ magnitudes, as well as the RGB tip $I$
magnitude obtained from theoretical isochrones 
are basically unaffected by the use of 
the appropriate $\alpha-$enhanced transformations.

We have also tested for the first time the effect of boundary
conditions obtained from appropriate $\alpha$-enhanced model
atmospheres on the stellar evolutionary tracks in the
log$L/L_{\odot}$-$T_{eff}$ plane. We find that, as in case of scaled
solar models, the integration of a solar $T(\tau)$ relationship
provides -- at least for masses larger than 0.5 -- 0.6 $M_{\odot}$ -- 
$\alpha$-enhanced tracks very similar to the ones computed
using boundary conditions from the appropriate model atmospheres.

\acknowledgments{S.C.\ and A.P.\ have been supported by MURST (PRIN2002, PRIN2003).
S.C.\ warmly acknowledges the hospitality at the Instituto de Astrofisica de
Canarias in Tenerife, and in particular A. Aparicio and C. Gallart for their
scientific as well as -- more important -- human support. We warmly thank our
anonymous second referee for his/her remarks that significantly improved this paper.}   
   

\clearpage   
\begin{deluxetable}{lcccccc}
\tablewidth{0pt}
\tablecaption{Colors predicted by model atmospheres computed for
a scaled solar mixture, a fully $\alpha-$enhanced one, and by changing the abundance of each individual $\alpha$ 
element at constant [Fe/H] (see text for details). \label{tab1}} 
\tablehead{
\colhead{Color}                       & 
\colhead{scaled solar\tablenotemark{a}} &
\colhead{$\alpha-$enhanced\tablenotemark{b}} &
\colhead{$Mg_{+0.4}$\tablenotemark{c}}       & 
\colhead{$O_{+0.4}$\tablenotemark{d}}   & 
\colhead{$Si_{+0.4}$\tablenotemark{e}}      &
\colhead{$Ca_{+0.4}$\tablenotemark{f}} 
}
\startdata
 $(B-V)$  & 0.841 & 0.793 & 0.799 & 0.813 & 0.839 & 0.850 \\
 $(U-B)$  & 0.322 & 0.231 & 0.253 & 0.314 & 0.284 & 0.324 \\ 
\enddata
\tablenotetext{a}{Scaled solar model atmosphere.}
\tablenotetext{b}{$\alpha-$enhanced model atmosphere.}
\tablenotetext{c}{Model atmosphere computed by enhancing by 0.4 dex the abundance of Mg (see text for details).}
\tablenotetext{d}{Model atmosphere computed by enhancing by 0.4 dex the abundance of O.}
\tablenotetext{e}{Model atmosphere computed by enhancing by 0.4 dex the abundance of Si.}
\tablenotetext{f}{Model atmosphere computed by enhancing by 0.4 dex the abundance of Ca.}
\end{deluxetable}

\clearpage   
\begin{deluxetable}{lccccc}
\tablewidth{0pt}
\tablecaption{Selected properties of 12~Gyr old $\alpha-$enhanced isochrones 
with the labeled metallicities; the table displays 
differences (in mag) between the values obtained using the 
the appropriate $\alpha-$enhanced transformations, and scaled-solar
ones with either the same [M/H] or the same
[Fe/H] (see text for details). \label{tab2}} 
\tablehead{
\colhead{CT transformations}                       & 
\colhead{$(B-V)_{\rm M_{\sl V}=6}$\tablenotemark{a}} &
\colhead{M$_{V}$(TO)\tablenotemark{b}}       & 
\colhead{$\Delta(B-V)$\tablenotemark{c}}   & 
\colhead{M$_{I}$(TRGB)\tablenotemark{d}}      &
\colhead{M$_{V}$(ZAHB)\tablenotemark{e}} 
}
\startdata
\multicolumn{6}{c}{Z=0.001}\\
 scaled-solar (same [M/H])     & $-$0.027 & 0.014 & $-$0.031 & 0.003 &    0.009 \\
 scaled-solar (same [Fe/H])    & $-$0.006 & 0.030 & $-$0.012 & 0.008 & $-$0.001 \\ 
\multicolumn{6}{c}{Z=0.004}\\ 
 scaled-solar (same [M/H])     & $-$0.052 &    0.008 & $-$0.042 & 0.002    &    0.009 \\
 scaled-solar (same [Fe/H])    & $-$0.028 & $-$0.014 & $-$0.027 & $-$0.001 & $-$0.002 \\ 
\multicolumn{6}{c}{Z=0.01}\\ 
 scaled-solar (same [M/H])     & $-$0.072 & $-$0.019 & $-$0.044 & 0.001 &    0.012 \\
 scaled-solar (same [Fe/H])    & $-$0.041 &    0.003 & $-$0.028 & 0.023 & $-$0.004 \\ 
\enddata
\tablenotetext{a}{$(B-V)$ color at M$_{V}=6.0$ mag.}
\tablenotetext{b}{Absolute visual magnitude of the Turn Off.}
\tablenotetext{c}{$(B-V)$ difference between the Turn Off and the RGB (see text for details).}
\tablenotetext{d}{Absolute $I$-Cousins magnitude of the RGB tip.}
\tablenotetext{e}{Absolute visual magnitude of the ZAHB at the level of the
RR Lyrae instability strip.} 
\end{deluxetable} 

\clearpage

\clearpage

\clearpage

\clearpage

\clearpage

\end{document}